# Strong exciton-plasmon coupling in MoS₂ coupled with plasmonic lattice

Wenjing Liu[1]‡, Bumsu Lee[1]‡, Carl H. Naylor[2], Ho-Seok Ee[1], Joohee Park[1], A.T. Charlie Johnson[1,2], and Ritesh Agarwal[1]*

Department of Materials Science and Engineering[1] and Department of Physics and Astronomy[2], University of Pennsylvania, Philadelphia, Pennsylvania 19104, USA

**ABSTRACT**: We demonstrate strong exciton-plasmon coupling in silver nanodisk arrays integrated with monolayer MoS₂ via angle-resolved reflectance microscopy spectra of the coupled system. Strong exciton-plasmon coupling is observed with the exciton-plasmon coupling strength up to 58 meV at 77 K, which also survives at room temperature. The strong coupling involves three types of resonances: MoS₂ excitons, localized surface plasmon resonances (LSPRs) of individual silver nanodisks and plasmonic lattice resonances of the nanodisk array. We show that the exciton-plasmon coupling strength, polariton composition and dispersion can be effectively engineered by tuning the geometry of the plasmonic lattice, which makes the system promising for realizing novel two-dimensional plasmonic polaritonic devices.

Few-layered transition metal dichalcogenides (TMDs) have recently attracted significant attention owing to their unique optical[1-3] and electronic[4, 5] properties. In particular, when thinned down to a monolayer, these materials become direct band gap semiconductors[6, 7] with large exciton binding energies[3], which makes them excellent candidates for achieving strong light-matter coupling[8, 9] for various applications including ultrathin, flexible devices. Strong coupling is achieved when the rate of energy transfer between the exciton and light is faster than their average dissipation rate, with the formation of light-matter hybrid states and new quasi-particles named exciton polaritons. As half-light half-matter particles, polaritons not only have the advantages of photons, such as small effective mass, fast propagation and long range spatial and temporal coherence, but also, via their matter fraction, possess strong inter-particle interactions that greatly enhances their nonlinear properties. This unique property allows for the observation of intriguing coherent phenomena such as polariton Bose-Einstein condensation[10, 11] and polariton bistability[12-14] as well as enables novel photonic devices such as polariton lasers[10, 15] and optoelectronic/all-optical circuit elements like polariton switches[16], transistors[12], and logic gates[13, 14, 16]. Moreover, the dispersion of polaritons can be significantly different from that of photons, which can generate slow light that is advantageous for sensing[17] and enhancing nonlinear processes[18]. Particularly, polaritons in 2-D systems with broken translational symmetry, have enhanced light-matter coupling strength due to the confinement of the excitons and the light field, and also enables exotic effect like superfluidity[19] and emerging of topological polariton states[20]. Understanding and tailoring the coupling strength, composition and dispersion of the polaritons is critical to the realization of these effects and the development of 2D polaritonic devices.

Recently, there has been a growing interest in studying strong light-matter coupling in exciton-plasmonic systems including metallic nanoparticles[21], gratings[22] and lattices[23-26]. Surface plasmon polaritons (SPPs) can tightly confine light much below the diffraction limit, which leads to a strongly enhanced local field with ultrasmall mode volume and hence enables strong exciton-plasmon coupling with observation of giant Rabi splitting of up to several hundred meV[22-26]. Among different plasmonic systems, plasmonic lattices are of special interest for strong coupling studies. Their building blocks, i.e. individual metallic nanostructures, can support localized surface plasmon resonances (LSPRs) with highly confined electric field in their vicinity, and when arranged into periodic arrays, these LSPRs can couple coherently to the different diffractive orders of the array and display significantly narrow lattice resonances via Rayleigh's anomaly[27, 28]. As a result, plasmonic lattices can combine the advantages of the enhanced local field of LSPRs and the higher quality factor of the coupled lattice resonances, therefore serving as good platforms for engineering strong exciton-plasmon coupling in excitonic materials. Moreover, in contrast to the resonant structure of the conventional dielectric cavities such as distributed Bragg reflectors (DBRs), in the plasmonic lattices, the two types of resonances with different physical properties can coexist and be tuned independently by changing the nanoparticle and lattice geometries, therefore allowing for greater freedom for tailoring the polariton composition and dispersion for engineering different properties in the system.

In spite of the great progresses in understanding the strong coupling phenomenon, practical application of polaritonic devices is hindered by the limitation of the properties of traditional materials. The small exciton binding energies of conventional semiconductors like GaAs preclude the observation of polaritonic effects at room temperature, as is also the challenging in highly lossy plasmonic systems. Organic materials, on the other hand, have large exciton binding energies and are thus mostly applied to study strong coupling in plasmonic systems, but suffer from disorder (inhomogeneous broadening) and photo-instability. At this stage, exploration of new exitonic materials suitable for practical devices is desired. TMDs have large exciton binding energies[3] due to the 2D quantum confinement of the excitons[29] as well as the reduced dielectric screening[30], which may overcome the high loss of the plasmonic systems and enable the robust exciton-plasmon coupling that is required for high temperature or high carrier density applications. Moreover, as crystalline semiconductors, they have significantly less inhomogeneous broadening and are highly photostable, and hence can be utilized for practical devices. Finally, atomically thin TMDs coupled to planar plasmonic lattices with thickness as small as tens of nanometers may pave the way for ultrathin and flexible polaritonic devices.

In our earlier work, we reported bowtie arrays coupled to monolayer $MoS_2$, which in the weak coupling regime exhibited large Purcell enhancement at room temperature and in the intermediate coupling regime demonstrated Fano resonances due to the exciton-plasmon interaction at 77 K[31]. By a careful design of geometrical parameters of the plasmonic lattices and characterization of the resonance modes of the system dispersions by Fourier optics, in this letter, we demonstrate for the first time the observation of strong coupling between the excitons in $MoS_2$ and plasmons at both cryogenic and room temperatures. We show that the coupling involves three types of resonances: excitons, lattice resonances, and LSPRs, which can be effectively modulated by tuning the geometries of the plasmonic lattices to produce polaritons with different exciton/plasmon compositions and hence dispersions.

$MoS_2$ samples used in our experiments were grown by chemical vapor deposition (CVD) on Si substrates with 275 nm thermally grown $SiO_2$.[32] Silver nanodisks arrays with varying diameters and pitch (square, two-dimensional lattice) were then fabricated on both MoS$_2$ monolayers and bare Si/SiO$_2$ substrates via electron-beam lithography (See supporting information), as shown in the SEM image in Figure 1 (a).[31] A home-built angle-resolved system was applied to measure the reflectance spectra of the samples (Figure 1 (b)), as demonstrated in detail in Ref. 33 (See supporting information). Nanodisks were used in this work instead of bowties for two reasons: first, although exceptionally strong field enhancement can occur between the two apexes of the bowtie resonator, the precise fabrication of a homogeneous bowtie array is difficult due to the small gap width between the two apexes (~10-20 nm). In contrast, nanodisk arrays with a simpler cylindrical geometry and diameters of the order of 100 nm are more tolerant to the fabrication errors, which is advantageous for observing strong coupling by reducing the inhomogeneous broadening. Secondly, by varying their diameters, the LSPRs of nanodisks can be more easily and precisely tuned to be in resonance with MoS$_2$ excitons (A and B) to enable a systematic study of the effect of coupling of LSPR resonances with excitons and the resulting polariton dispersions.

In order to understand the plasmonic resonances of the silver nanodisk array system with varying diameter and pitch, we first studied silver nanoparticle arrays on Si/SiO$_2$ substrates without MoS$_2$. LSPRs of individual silver nanodisks with diameters ranging from 90-150 nm and 50 nm thickness were first studied by measuring the far-field reflectance of fabricated arrays with lattice constant of 1000 nm (See supporting information). The interaction between nanodisks is weak in these lattices due to the large lattice constant, and measured dispersions via angle-resolved reflectance microscopy further confirmed that no lattice resonances were observed within 550 nm - 700 nm, the wavelength range relevant to the excitonic region of MoS$_2$ (See supporting information). Therefore, the measured resonances of these lattices can be approximately assumed to be the LSPRs of individual nanodisks. A dip in the differential reflectance ($(R - R_0)/R_0$) spectra of the arrays was observed and found to redshift with increasing disk diameters, as shown in Figure 1 (c). To understand the measured data, Finite-difference time-domain (FDTD) simulations were performed to calculate the LSPR wavelengths of single silver nanodisk on $Si/SiO_2$ substrate with different diameters. We found that the simulated wavelengths of the first order plasmonic whispering gallery mode matched well with the dip positions in the measured reflectance spectra (Figure 1 (d)). Therefore the reflectance dips can be attributed to the first order LSPR modes of the individual nanodisks with dipolar field distributions (Figure 1 (d), inset) that allow efficient far-field coupling.

After characterizing the LSPR of individual Ag nanodisks, we then reduced the lattice constant of the arrays to enable the interaction between the nanodisks. The five different fabricated arrays contained $100 \times 100$ equally sized silver nanodisks arranged in a square lattice with the lattice constants of all arrays fixed to 460 nm, while the diameters of the nanodisks in different samples varied from 100 nm to 170 nm to allow their LSPR modes to span the 590 nm - 650 nm wavelength range, the excitonic region of $MoS_2$. Sharp lattice resonances were subsequently observed due to the collective diffraction of the array in the measured angle-resolved reflectance spectra (Figures 1 (e)-(j)) for all the five arrays patterned on a $Si/SiO_2$ substrate. For all samples, the lowest energy lattice resonances, i.e., the (±1,0) diffractive orders were observed with the Γ point (in plane wavevector $k_{\parallel}$ = 0) near 560 nm. Significantly, as observed in Figure 1 (f)-(h), for arrays with average disk diameters of 120 nm, 135 nm and 150 nm, we observed a pronounced anti-crossing of the dispersion curve near the LSPR wavelengths (white dashed lines), which indicates strong coupling between the LSPR and the lattice resonances. Such strong coupling phenomenon has also been observed in previous reports in plasmonic gratings[34] and lattices[23, 35], implying that the localized resonance of individual nanostructures can be coherently coupled to the collective diffraction of the lattices in these systems.

The dispersion curves were then fitted to a coupled-oscillator model (COM), with the Hamiltonian given by:

$$H = \begin{bmatrix} E_{S+} - i\gamma_{S+} & g_{\pm} & g_{SL} \\ g_{\pm} & E_{S-} - i\gamma_{S-} & g_{SL} \\ g_{SL} & g_{SL} & E_{LSPR} - i\gamma_{LSPR} \end{bmatrix}$$

where for the diagonal terms, $E$ and $\gamma$ denote the energy and the damping (half-width at half-maximum[35]) of each mode and the subscripts $S_+$, $S_-$ and LSPR stand for (+1,0), (-1,0) diffractive order and LSPR, respectively. The off diagonal term, $g$, represents the coupling strength between the two resonances. For simplicity, the arrays are assumed to be symmetric for positive and negative in-plane wavevector $k_{||}$, hence the coupling strengths of LSPR to both the (+1,0) and (-1,0) diffractive orders were set to be the same as $g_{SL}$. The system eigenstates were solved and fitted to the experimental data, with the linear $E$-$k$ dispersions of (±1,0) diffractive orders obtained from the dispersions of array with detuned LSPR (Figure 1 (j), d = 170 nm), and LSPR energies and coupling strengths used as fitting parameters. The fitted LSPR energies for all disk sizes agree well with both experimental data and FDTD simulation results for individual (uncoupled) nanodisks as shown in Figure 1 (d). Coupling strengths between LSPR and diffractive modes obtained from the COM are 95 meV, 110 meV and 125 meV, for arrays with average disk diameters of 120 nm, 135 nm and 150 nm, respectively. The lattice-LSPR coupling strength is positively correlated with the disk diameter likely due to the increase of the nanodisk polarizability with increasing diameter. The coupling strength between (+1,0) and (-1,0) diffractive orders, $g_\pm$, was found to be negligible in all the measured arrays. The three fitted polariton branches with negative $k_\parallel$ (Figure 1 (f)-(h), blue solid curves) show the dispersions and quality factors of the coupled lattice-LSPR modes change significantly as a function of lattice resonance-LSPR detuning. Away from the LSPR wavelengths, the coupled mode is dominated by the lattice resonance and therefore has a higher quality factor and a steeper dispersion, while near LSPR wavelengths, it becomes more LSPR-like with a lower quality factor and a nearly flat dispersion.

After characterizing the plasmonic resonances of the Ag nanodisk arrays on $Si/SiO_2$, we investigated the properties of the plasmonic nanodisk array-coupled $MoS_2$ monolayer system. The experiment was first carried out at 77 K to reduce the exciton damping, while lowered temperature has little effect on plasmonic resonances. For arrays patterned on $MoS_2$, the LSPR modes redshifts by ~50-60 nm due to the higher refractive index of $MoS_2$[36], which is also confirmed by reflectance measurements on uncoupled arrays, with and without $MoS_2$ (supporting information). In Figure 2 (a)-(d), the angle-resolved differential reflectance spectra of $MoS_2$ coupled with four different arrays are presented. Each array was designed with the same lattice constant of 460 nm but with different nanodisk diameters so that the LSPR wavelengths can be tuned across the excitonic region of $MoS_2$ (590 nm - 650 nm). Bare monolayer $MoS_2$ displays two reflection dips corresponding to A- and B- excitons at ~640 nm and 590 nm respectively, arising from its direct band gap and strong spin-orbit coupling characteristics.[6, 7] For all nanodisk array coupled to $MoS_2$ samples, anti-crossing of dispersion curves near both A and B excitons were observed, indicating strong exciton-plasmon coupling. The dispersion curves for all four different arrays were fitted to a coupled oscillator model (COM, as described earlier) including five oscillators: A and B excitons, (+1,0) and (-1,0) diffractive orders (modes), and nanodisk

LSPR. In the COM fitting, the energies of A and B excitons as measured from the reflectance spectra from the same $MoS_2$ flake in the region without the nanodisk array were used; both lattice resonance dispersions and LSPR energies were treated as fitting parameters to account for the mode redshift caused by the presence of $MoS_2$, with the dispersions of the lattice resonances kept the same for all different lattices as they are solely determined by the lattice constants; coupling strengths between the A and B excitons, and between the two diffractive orders were set to zero, while other coupling strengths were used as fitting parameters. The values for the LSPR wavelengths and coupling strengths as obtained from the COM fitting are listed in Table 1. The lattice-LSPR coupling strength is of the order of 100 meV and increases with increasing disk diameter, which is consistent with the observations from the arrays on bare $Si/SiO_2$ substrate without $MoS_2$. The exciton-LSPR coupling is enhanced strongly with decreasing exciton-LSPR detuning, and the maximum coupling strength of 58 meV was found for the A exciton-LSPR coupling when the LSPR is nearly in resonance with the A exciton (d = 100 nm, LSPR ~630 nm). The exciton-lattice diffraction mode coupling was found to be the weakest among the three types of coupling, with the coupling strengths of the order of 10-20 meV. Therefore, in this $MoS_2$-plasmonic lattice system, the exciton-plasmon strong coupling is mainly mediated by the LSPR which strongly couples to both A and B excitons and the lattice resonances[23]. Although LSPRs are highly localized, the presence of long-range lattice diffraction modes leads to coherent coupling of the $MoS_2$ exciton-LPSR "plexitonic" system at different regions, much beyond the lengthscales of the localized plasmons or excitons thereby increasing the oscillator strength of various coupled resonances.

The coupling between the five different resonances results in five polariton branches with different exciton and plasmonic contributions and hence dispersions (Fig 2). Since for all the four arrays, the highest energy polariton branch lies far above the excitonic region and does not show up in the wavelength range of the measurement, we will focus our discussion on the remaining four branches with lower energies. The fitting results of these four branches are plotted for negative $k_∥$ in Figure 2 (a)-(d) (blue solid lines) and labeled as 1-4 from high to low energies. To further understand the strong exciton-plasmon coupling behavior of the system, the fractions of excitons and plasmons in the four polariton branches of different $k_∥$ were also calculated and plotted for the following three cases (Figure 3): LSPR energy larger than the B exciton (d = 70 nm, see Figure 2 (a)), LSPR energy between A and B excitons (d = 100 nm, see Figure 2 (b)), and LSPR energy lower than the A exciton (d = 120 nm, see Figure 2 (c)). Interestingly, the experimental and fitting results for arrays with different disk diameters illustrate that the strong exciton-plasmon coupling behavior of the system, including the polariton composition, dispersion, and coupling strength, can all be effectively tailored by changing the geometrical factors of the lattice, in this case the nanodisk diameter. First, the exciton and plasmon fractions of each polariton branches can be controlled by the relative spectral positions of LSPR and excitons. For the A exciton, when the LSPR is positively detuned from the A exciton (d = 70 nm, Figure 3 (a)), the A exciton is mainly distributed in branch 3 with ~0.9 exciton fraction at high $k_∥$, and in branch 4 with ~0.8 exciton fraction at low $k_∥$. As the LSPR energy redshifts (d = 100 nm, Figure 3 (b)) and becomes lower than the A exciton (d = 120 nm, Figure 3(c)), the A exciton component gradually transfers into branch 2 with ~0.8 exciton fraction at high $k_∥$ and branch 3 with ~0.9 exciton fraction at low $k_∥$. For the B exciton on the other hand, while present in all four lattices, the proportion of the B exciton is mainly distributed in polariton branches 1 and 2, and as the LSPR-B exciton detuning changes from positive (d = 70 nm) to negative (d = 100 nm), the relative fractions of the B exciton and the LSPR in these two branches at high $k_\parallel$ reverses from ~0.7 B exciton and ~0.3 LSPR to ~0.05 B exciton and ~0.95 LSPR (Figure 3 (d)-(f)). Secondly, the dispersions of the polaritons changes drastically for different lattice designs. Among the three types of resonances, LSPRs have flat dispersions as they do not propagate; the effective masses of excitons are orders of magnitude larger than the lattice modes and hence also show nearly flat dispersions; therefore, the polariton dispersion is mainly determined by the fraction of the lattice diffraction resonances. For instance, in polariton branch 2, the dispersion is almost flat in the lattice with d = 70 nm where its lattice mode fraction is less than 0.2 for all $k_\parallel$ (Figure 3 (g)). As the LSPR redshifts with increasing disk diameters, the LSPR fraction in this polariton branch decreases and the fraction of the lattice resonances increases, and hence the dispersion of this polariton branch becomes steeper. Finally, as discussed earlier, the exciton-plasmon coupling strength can be strongly enhanced when the LSPR mode is in resonance with the excitons, as shown in Table 1. In many cases, this change of coupling strength can also be characterized by the amount of splitting between the two polariton branches surrounding the excitons. For example, for the B exciton-plasmon coupling, when the LSPR is tuned close to the B exciton resonance, the splitting between polariton branch 1 and 2 is as large as 70 meV (d = 70 nm, Figure 2 (a)). It decreases when the LSPR is gradually detuned from the B exciton, and finally reduces to 10 meV for d = 140 nm, corresponding to a large LSPR-B exciton detuning of 320 meV.

The above results demonstrate that the tunable mode structure and distinct properties of different resonances in a plasmonic lattice allow engineering of the exciton-plasmon coupling strength, polariton composition, and therefore their physical properties, which is crucial for polariton devices design tailored for specific applications. Specifically, the large exciton fraction increases the polariton lifetimes and enables stronger inter-particle interaction, the large LSPR fraction enhances the coupling strength, and the lattice resonance fraction tailors the polariton dispersions, and its collective nature enable the coherent coupling for excitons within its coherence length (usually >10X lattice parameter). Proper tuning of these compositions by carefully designing the plasmonic lattices will allow for the development of polaritonic devices that are suitable for a variety of purposes.

Finally, to evaluate the robustness of the strong exciton-plasmon coupling of the system, we studied its temperature dependence by measuring the reflectance spectra for both silver array-coupled $MoS_2$ (d = 120 nm, lattice constant 460 nm) and bare $MoS_2$ monolayer as a function of temperature. For silver array-coupled $MoS_2$, anti-crossing of the polariton dispersions at room temperature is not as evident as it is a 77 K (Figure 4 (a) and (b)), which is due to increased exciton damping at high temperature indicated by the linewidth broadening of both A and B excitons (Figure 4 (d)). However, near the A exciton region (~660 nm), strong coupling can still be resolved in the angle-selected reflectance spectra of $MoS_2$ coupled to silver nanodisk array (Figure 4c). As seen in Figure 4c, the reflectance dips for polariton branches 2 and 3 can be identified at the high- and low-energy sides of the A exciton, respectively. This finding demonstrates that strong exciton-plasmon coupling survives at room temperature, probably owning to the large exciton binding energy of $MoS_2$ that is at least one order of magnitude larger than the thermal energy at the room temperature (~26 meV). Fitting these two branches to the coupled oscillator model with exciton energies and linewidths at room temperature gives an A exciton-LSPR coupling strength of 43 meV and an A exciton-lattice resonance coupling strength of 11 meV, which is decreased only slightly from the values at 77 K. This result indicates that strong coupling is still robust at room temperature, and hence the system holds promise for room temperature plasmonic polaritonic devices in ultrathin semiconductors. It is also worth noting that this coupling strength value is much larger than the exciton-photon coupling strength of 23 meV reported in $MoS_2$ coupled with dielectric cavities at room temperature[8], and therefore highlights the strongly enhanced light-matter interaction strengths in plasmonic lattices.

For a more detailed study of the temperature dependence of exciton-plasmon polaritons, the wavelengths of the A exciton and the two polariton branches at $\sin\theta = -0.37$ (white dashed vertical line in Figure 4 (a)) as a function of temperature were extracted (Figure 4 (f)). At low temperatures, the energy of polariton branch 2 at the selected angle is closer to the A exciton than polariton branch 3, and it also redshifts faster with increasing temperature. While at high temperatures, the rate of redshift of polariton branch 3 increases as its energy approaches the A exciton where the energy of branch 2 becomes nearly temperature independent. The reason for this phenomenon is that the redshift of the A exciton changes the exciton-LSPR detuning, and therefore changes the relative fractions of the exciton and plasmon of the polaritons. Quantitatively, as shown in the calculation results of the exciton fractions in Figure 4 (g), at the selected angle, polariton branch 2 contains 0.7 exciton fraction at 77 K and is more matter-like and temperature dependent, while at the room temperature, this proportion drops to 0.3, and it becomes more light-like and temperature independent. For polariton branch 3, on the other hand, the exciton fraction increases from 0.35 to 0.75 from 77 K to room temperature, therefore its temperature dependence shows the opposite trend. This result further illustrates that the physical properties, in this case the temperature dependence, of the light-matter hybrids in emitter-coupled plasmonic lattices can be tailored by tuning the relative fractions of different resonances with different features, which will be helpful for designing polaritonic devices with precisely tailored responses.

To conclude, we have observed strong exciton-plasmon coupling in silver nanodisk arrays integrated with monolayer $MoS_2$. Strong exciton-plasmon coupling was studied by angle-resolved reflectance measurement, FDTD simulations, and the coupled oscillator model. We show, for the first time, the evidence of strong coupling between surface plasmon polaritons and $MoS_2$ excitons involving three types of resonances: excitons, plasmonic lattice resonances and localized surface plasmon resonances, with the ability to effectively tune the coupling strength and physical properties of plasmon-exciton polaritons by changing the diameter of the nanodisks. By combining the distinct properties of LSPRs and long range diffraction modes of the plasmonic lattices and the unique properties of 2D semiconductors such as their large exciton binding energies and atomic thickness, unexplored coherent phenomena like plasmonic polariton Bose-Einstein condensation as well as ultrathin and robust plasmonic polariton devices may become a reality.

**FIGURE CAPTIONS**

**Figure 1. Sample configuration and the optical properties of the plasmonic lattice patterned on bare Si/SiO₂ substrates.** (a) SEM image of a typical silver nanodisk lattice patterned on monolayer $MoS_2$. (b) Schematic of the angle-resolved reflectance measurement setup[33]. (c) Far-field differential reflectance of six plasmonic lattices with nanodisk diameters ranging from 90-154 nm and 1000 nm lattice constant (uncoupled nanodisks). (d) LSPR wavelengths for the first and second order whispering gallery LSPR modes in individual silver nanodisks obtained by FDTD simulations along with the experimentally measured the dip positions from the reflectance spectra in (c). Insets show the simulation configuration and the z-polarized E-field distribution in the x-y plane for the first and second order whispering gallery LSPR modes in silver nanodisks. (e)-(i) The angle-resolved differential reflectance spectra for five plasmonic lattices patterned on bare Si/SiO₂ substrates with disk diameters ranging from 100-170 nm and 460 nm lattice constant. White dashed line represents the wavelength of the LSPR modes and the red dots correspond to the dip positions obtained from the line cuts of the angle-resolved reflectance at constant angles. Blue solid lines are the fitting results obtained from a coupled-oscillator model.

**Figure 2. Strong exciton-plasmon coupling of silver plasmonic lattice-coupled monolayer MoS₂ at 77K.** (a)-(d) Angle-resolved differential reflectance spectra of four different silver nanodisk arrays with different disk diameters patterned on monolayer $MoS_2$. The lattice constants of all four arrays are fixed at 460 nm while the disk diameters are changed. White

dashed lines correspond to the LSPR positions and yellow dashed lines represent the A (low energy) and B (high energy) excitons. Red dots correspond to the dip positions obtained from the line cuts of the angle-resolved reflectance at constant angles. Blue solid lines are the fitting results from a coupled-oscillator model. (e)-(h) Line cuts from the angle-resolved reflectance data taken from a constant angle represented by the vertical white dotted lines in (a)-(d), respectively. The dip positions are indicated by the blue triangles.

**Figure 3. Composition of the four polariton branches for three representative LSPR-exciton detuning conditions in silver plasmonic lattice-coupled monolayer MoS$_2$ at 77K.** (a)-(c) A exciton fraction. (d)-(f) B exciton fraction. (g)-(i) lattice mode fraction. (j)-(l) LSPR fraction.

**Figure 4. Temperature dependence of the strong exciton-plasmon coupling in silver plasmonic lattice-coupled monolayer MoS$_2$** (a)-(b) Angle-resolved differential reflectance spectra of the silver nanodisk array coupled with MoS$_2$ at 77 K and 300 K, respectively. The disk diameter is 120 nm with a lattice constant of 460 nm. (c) The spectral line cuts taken from room temperature data in (b) with the incident angles θ ranging from 21.7$^o$ to 42.1$^o$ (d) Temperature dependent differential reflectance spectra of bare MoS$_2$. (e) Temperature dependent differential reflectance spectra of silver nanodisk array coupled with MoS$_2$ obtained from the angle-resolved differential reflectance spectra at $\sin \theta = -0.37$. The position of the line cut is shown in figure (a) as a vertical white dotted line. (f) The wavelength of the A exciton and polariton branches 2 and 3 at $\sin \theta = -0.37$ as a function of temperature, extracted from the dip positions in (d) and (e). (g)

Exciton fractions of polariton branches 2 and 3 at $\sin\theta = -0.37$ as a function of temperature, calculated from the COM with the temperature dependent exciton energies extracted from (d).

**Table 1. Coupling constants obtained from fitting the experimentally measured polariton dispersion to a coupled oscillator model. All quantities are in units of meV.**

| Diameter | | A exciton | B exciton | LSPR |
|---|---|---|---|---|
| d = 72 nm LSPR: 590 nm | Lattice resonance | 9 | 13 | 90 |
| | LSPR | 45 | 43 | -- |
| d = 98 nm LSPR: 630 nm | Lattice resonance | 16 | 11 | 97 |
| | LSPR | 58 | 40 | -- |
| d = 121 nm LSPR: 660 nm | Lattice resonance | 17 | 16 | 116 |
| | LSPR | 44 | 26 | -- |
| d = 149 nm LSPR: 710 nm | Lattice resonance | 15 | 9 | 133 |
| | LSPR | 35 | 17 | -- |

ASSOCIATED CONTENT

**Supporting Information**. The methods of sample synthesis, device fabrication and angle-resolved measurement. Reflectance spectra of uncoupled nanodisk array (1000 nm pitch) fabricated on bare Si/SiO$_2$ substrate and on monolayer MoS$_2$. Angle-resolved reflectance spectra of nanodisk arrays with 120 nm diameter and different lattice constants. This material is available free of charge via the Internet at http://pubs.acs.org.

AUTHOR INFORMATION

**Corresponding Author**

E-mail: riteshag@seas.upenn.edu

**Author Contributions**

‡These authors contributed equally.

**Notes**

The authors declare no competing financial interest.

**ACKNOWLEDGEMENTS**

This work was supported by the NSF under the NSF 2-DARE program (EFMA-1542879, NSF-MRSEC (LRSM) seed grant under award number DMR11-20901 and by the US Army Research Office under Grant No. W911NF-11-1-0024. Nanofabrication and electron microscopy characterization was carried out at the Singh Center for Nanotechnology at the University of Pennsylvania.

**Figure 1.**

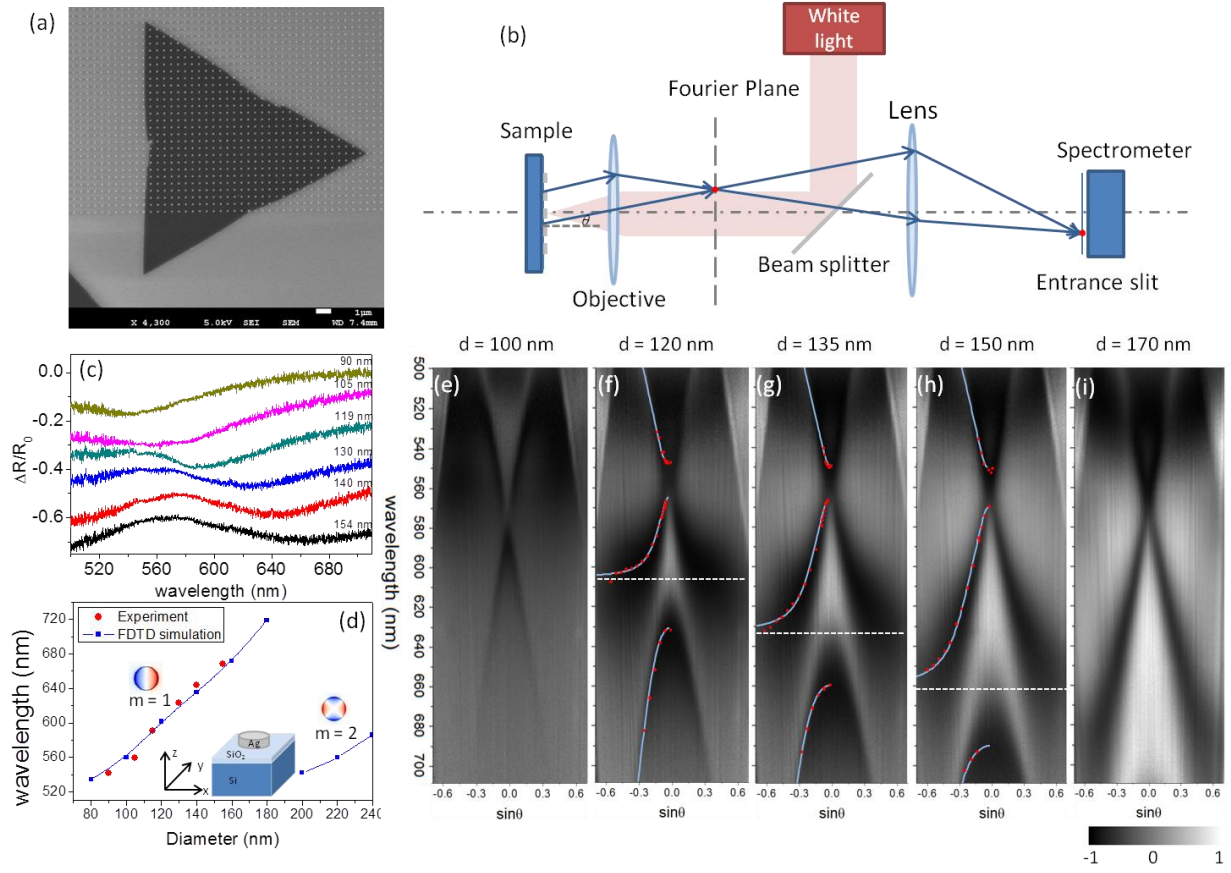

**Figure 2.**

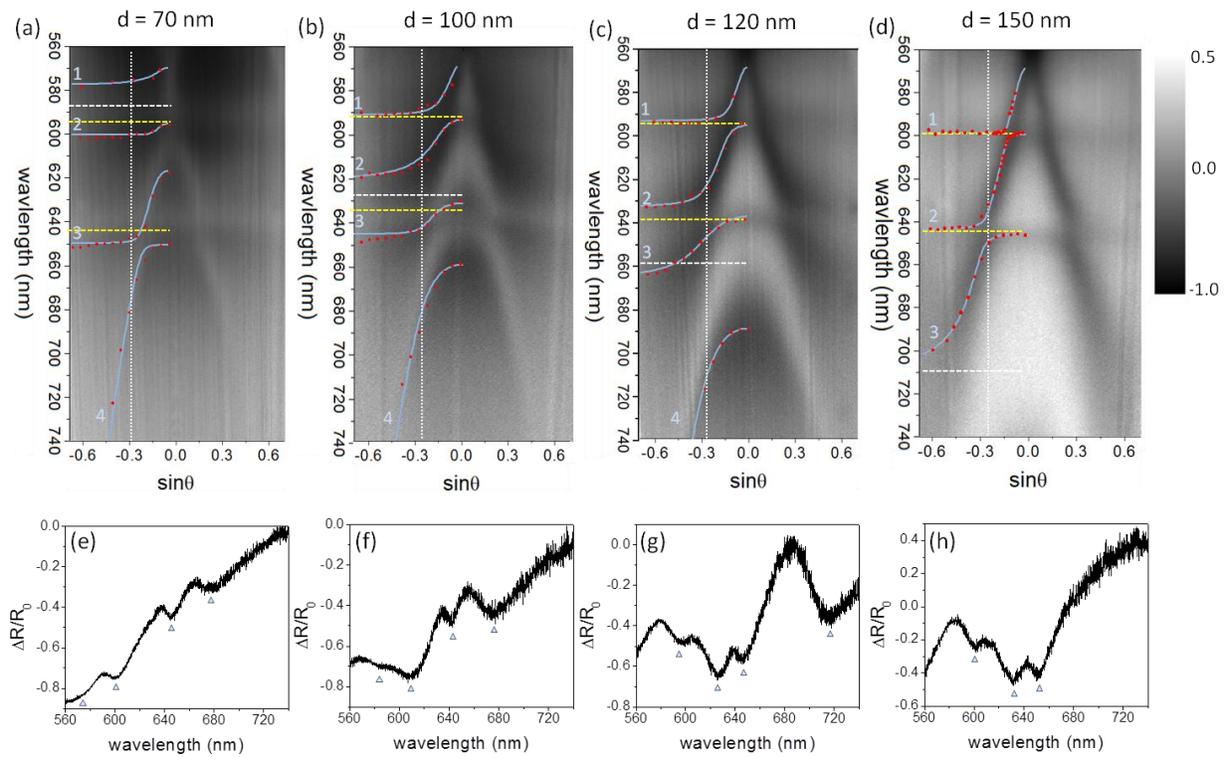

**Figure 3.**

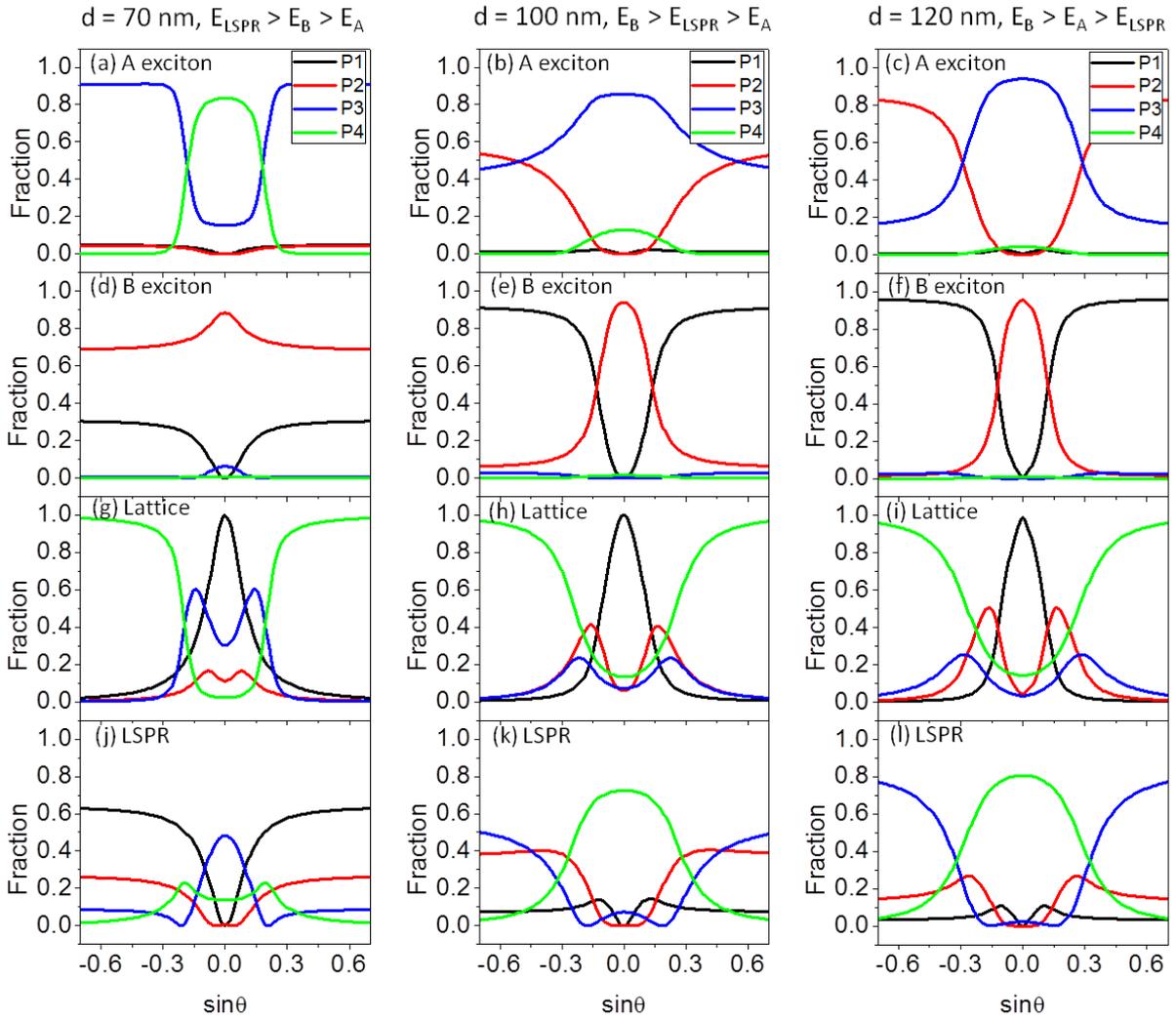

**Figure 4.**

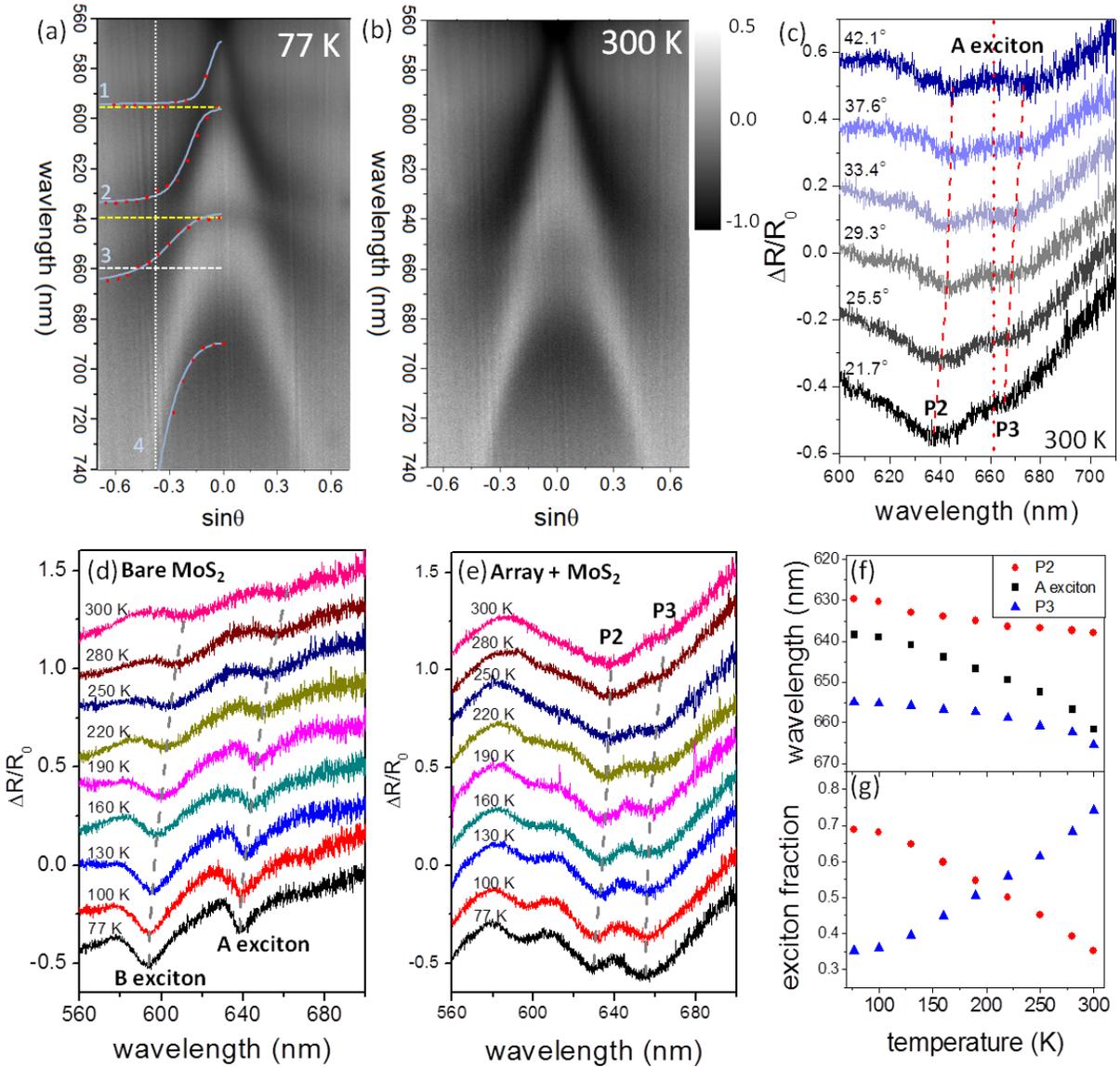

# Supporting information for strong exciton-plasmon coupling in MoS₂ coupled with plasmonic lattice

*Wenjing Liu[1]‡, Bumsu Lee[1]‡, Carl H. Naylor[2], Ho-Seok Ee[1], Joohee Park[1], A.T. Charlie Johnson[1,2], and Ritesh Agarwal[1]\**

Department of Materials Science and Engineering[1] and Department of Physics and Astronomy[2],

University of Pennsylvania, Philadelphia, Pennsylvania 19104, USA

E-mail: riteshag@seas.upenn.edu

**MoS₂ growth**: Single crystal MoS₂ flakes were grown directly on a 275nm Si/SiO₂ substrate by chemical vapor deposition[1]. A 1% sodium cholate solution was initially spin coated onto the SiO₂ substrate to help promote the growth region. A micro-droplet of a saturated solution of ammonium heptamolybate (AHM) was deposited onto a corner of the substrate, which acted as the molybdenum feedstock. The substrate was placed in the center of a 1-inch Lindberg blue furnace and 25 mg of solid sulfur (part number 213292, SigmaAldrich) was placed upstream at a distance of 18 cm from the growth substrate. 700 s.c.c.m of nitrogen was flown through the chamber and the temperature of the furnace was ramped up to 800 ℃, while the sulfur pellet was heated up to 150℃. After a 30 min growth, the furnace was then stopped and rapidly cooled to room temperature.

**Device fabrication**: A layer of PMMA 950 A2 was spin coated onto the as grown $MoS_2$ sample at 2000 r.p.m. for 45 s and baked at 180 ℃ for 90 s. Electron beam lithography was used to define the nanodisk arrays. 50 nm thick silver film was deposited by electron-beam deposition followed by a lift-off process. The nanodisk diameter and array pitch were measured by scanning electron microscopy (SEM).

**Angle-resolved reflectance measurement**: A home-built angle-resolved system[2] was used to measure the reflectance spectra of the sample, as shown in figure 1 (b). A white light beam was focused on the sample by a microscope objective (60×, NA = 0.7, Nikon). Parallel light reflected from the sample was focused at the back focal plane (Fourier plane) of the objective and a lens was used to project the Fourier plane onto the entrance slit of a spectrometer (Princeton Instruments). The spectrometer CCD (2048 × 512 pixels) recorded both the wavelength of light and its spatial position at the entrance slit (angles of the reflected light) to extract the angle- and wavelength- resolved reflectance spectrum.

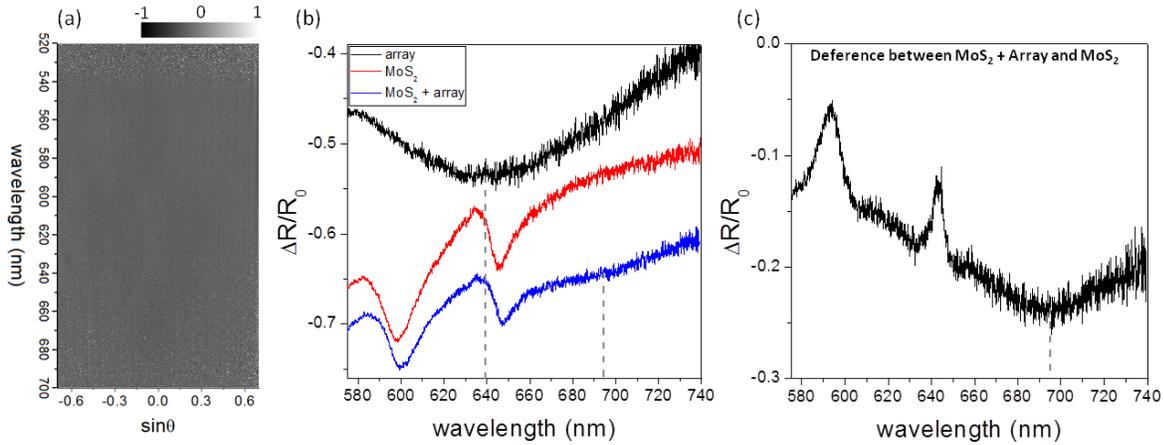

**Figure S1. LSPR modes of individual (uncoupled) nanodisks patterned on Si/SiO₂ substrate and on monolayer MoS₂ with disks diameter of d = 140 nm and lattice constant, a = 1000 nm.** (a) angle-resolved reflectance spectra of the Ag nanodisk array patterned on Si/SiO₂ substrate. No clear lattice resonances can be observed. (b) Far-field reflectance measurement of the Ag nanodisk array on Si/SiO₂ substrate and on monolayer MoS₂. The resonance position (reflectance dip) is redshifted from 640 nm to 695 nm due to the presence of MoS₂. (c) The difference of the reflectance spectra between the array patterned on MoS₂ and bare MoS₂ from (b). The LSPR position (reflectance dip) can be clearly observed at 695 nm.

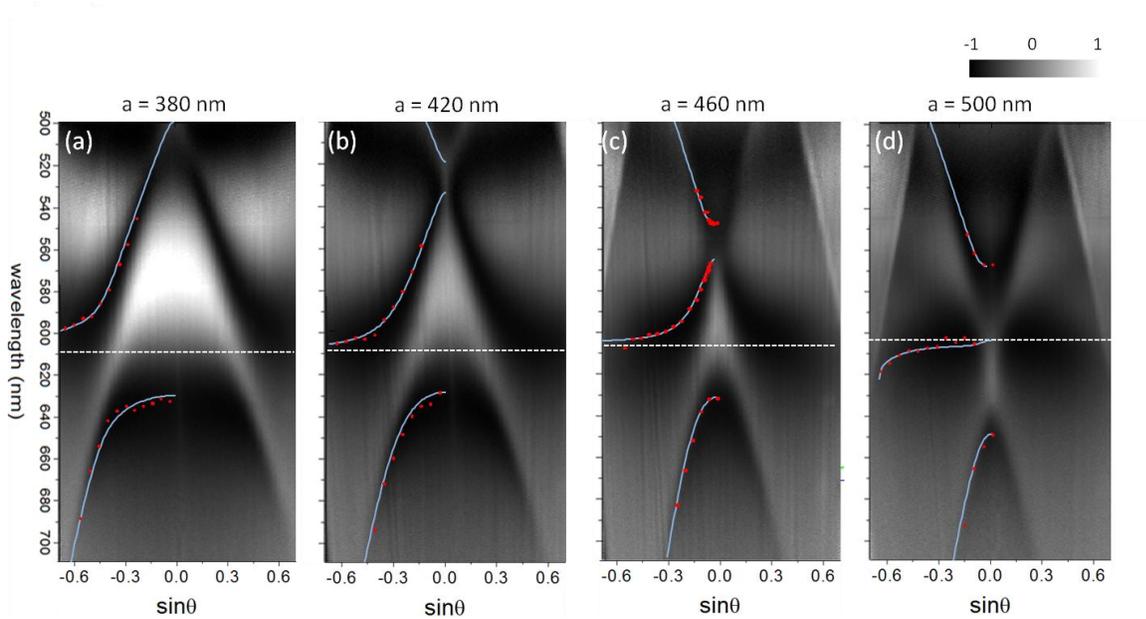

**Figure S2. Angle-resolved reflectance spectra of four different arrays with disks diameter fixed at d = 120 nm and for various lattice constants (*a*).** (a) a = 380 nm. (b) a = 420 nm. (c) a= 460 nm. (d) a = 500 nm. White dashed line represents the wavelength of the LSPR modes and the red dots correspond to the dip positions obtained from the line cuts of the angle-resolved reflectance at constant angles. Blue solid lines are the fitting results obtained from the coupled-

oscillator model. In (d), the second order diffractive modes have also been considered in the COM fitting.